# Crystal fields, disorder, and antiferromagnetic short-range order in $Yb_{0.24}Sn_{0.76}Ru$

T. Klimczuk[1,2], C. H. Wang[3,4], J. M. Lawrence[3], Q. Xu[5,6], T. Durakiewicz[1], F. Ronning[1], A. Llobet[1], F. Trouw[1], N. Kurita[1], Y. Tokiwa[1], Han-oh Lee[1], C. H. Booth[7], J. S. Gardner[8,9], E. D. Bauer[1], J. J. Joyce[1], H. W. Zandbergen[4], R. Movshovich[1], R. J. Cava[10], and J. D. Thompson[1]

[1] Los Alamos National Laboratory, Los Alamos, New Mexico 87545, USA

[2] European Commission, JRC, Institute for Transuranium Elements, Postfach 2340, 76125 Karlsruhe, Germany,

[3] University of California, Irvine, California 92697, USA

[4] Neutron Scattering Science Division, Oak Ridge National Laboratory, Oak Ridge, Tennessee 37831, USA

[5] National Centre for HREM, Department of Nanoscience, Delft Institute of Technology, 2628 CJ Delft, The Netherlands

[6] EMAT, University of Antwerp, 2020 Antwerp Groenenborgerlaan, 171,U316, Belgium

[7] Chemical Sciences Division, Lawrence Berkeley National Laboratory, One Cyclotron Rd., Berkeley, California 94720, USA

[8] NCNR, National Institute of Standards and Technology, Gaithersburg, Maryland 20899-6102, USA

[9] Indiana University, Bloomington, Indiana 47408, USA

[10] Department of Chemistry, Princeton University, Princeton NJ 08544, USA

Abstract

We report extensive measurements on a new compound $(Yb_{0.24}Sn_{0.76})Ru$ that crystallizes in the cubic CsCl structure. Valence band photoemission and $L_3$ x-ray absorption show no divalent component in the 4f configuration of Yb. Inelastic neutron scattering (INS) indicates that the eight-fold degenerate $J$-multiplet of $Yb^{3+}$ is split by the crystalline electric field (CEF) into a $\Gamma_7$ doublet ground state and a $\Gamma_8$ quartet at an excitation energy 20 meV. The magnetic susceptibility can be fit very well by this CEF scheme under the assumption that a $\Gamma_6$ excited state resides at 32 meV; however, the $\Gamma_8/\Gamma_6$ transition expected at 12 meV was not observed in the INS. The resistivity follows a Bloch-Grüneisen law shunted by a parallel resistor, as is typical of systems subject to phonon scattering with no apparent magnetic scattering. All of these properties can be understood as representing simple local moment behavior of the trivalent Yb ion. At 1 K, there is a peak in specific heat that is too broad to represent a magnetic phase transition, consistent with absence of magnetic reflections in neutron diffraction. On the other hand, this peak also is too narrow to represent the Kondo effect in the $\Gamma_7$ ground state doublet. On the basis of the field-dependence of the specific heat, we argue that antiferromagnetic short-range order (possibly co-existing with Kondo physics) occurs at low temperatures. The long-range magnetic order is suppressed because the Yb site occupancy is below the percolation threshold for this disordered compound.

**Introduction**

Often considered to be 4*f* hole analogs of Ce-based 4*f* electron heavy-fermion systems, Yb-based heavy-Fermion compounds exhibit a remarkable breadth of strongly correlated electron behavior that can be tuned readily by crystal chemistry and external control parameters. For example, $YbB_{12}$ is a Kondo insulator [1]; metallic $YbInCu_4$ exhibits a first-order valence transition, and the physical properties of isostructural $YbAgCu_4$ are well-described by the Anderson impurity model [2]. The unconventional superconductivity and quantum critical states that are found in Ce heavy-Fermion materials also appear in Yb systems. In contrast to the growing number of Ce-based heavy-Fermion superconductors, there is, so far, only one Yb-based example (β- $YbAlB_4$ [3]), but there are several stoichiometric Yb-compounds, such as $YbRh_2Si_2$ [4], YbAgGe [5], and YbPtIn [6], and alloys (Ir [7]-, Ge [8]- and La [9]-doped $YbRh_2Si_2$) for which a field-induced quantum critical point (QCP) has been reported. Clearly these Yb-based materials have yielded a rich variety of phenomena. As compared to the large number of known correlated Ce-based materials, however, relatively few Yb compounds have been explored. In part, this is due to the high volatility of Yb, which makes sample preparation more challenging.

In a search for new heavy-Fermion compounds, we have investigated the Yb-Ru-Sn ternary system and have uncovered a new compound with stoichiometry $Yb_{0.24}Sn_{0.76}Ru$. This compound, with the simple CsCl structure [10], is isostructural with the YbRu binary, for which no physical properties have been previously reported [11]. To explore the extent of heavy-Fermion physics in this compound, we undertook a series of measurements of the temperature dependence of the susceptibility and resistivity as well as the temperature and magnetic field dependence of the specific heat. To help interpret results of these measurements, we also determined the electronic structure and possible role of crystalline electric fields through valence band photoemission, $L_3$ x-ray absorption, and inelastic neutron scattering experiments. As we will show, these experiments are consistent with the presence of trivalent Yb local moments in a crystalline electric field of cubic site symmetry. On the basis of both neutron diffraction and the field dependence of the specific heat, we argue that antiferromagnetic short-range (and possibly also Kondo physics [12]) occurs at low temperatures in this compound; long-range magnetic order is suppressed by alloy disorder.

**Experimental details**

Single crystals of $Yb_{0.24}Sn_{0.76}Ru$ were grown from high temperature solution. The starting high purity (99.9% Ru, 99.99% Sn, 99.999% Pb and 99.9% Yb) metals in the ratio of Yb:Ru:Sn:Pb = 1.1:2:1:15 were placed in an alumina crucible and sealed under vacuum in a quartz tube. The tube was heated to 1150 °C and kept at that temperature for six hours, then cooled at the rate of 5 °C /hr to 650 °C, at which temperature the excess Pb was spun off with the aid of a centrifuge. The obtained crystals, in the shape of regular cubes, were etched in 1:1 $H_2O_2$ and acetic acid mixture to remove remaining Pb flux. The small size of these cubes (0.3×0.3×0.3 $mm^3$) is typical for Heusler systems; for example $MPd_2Pb$ (M=Sc, Y, Gd, Tb, Dy, Ho, Er, Tm and Lu) crystals grown in Pb flux reach the size of 1 mm on a side [13].

Characterization of the elemental composition was obtained by electron probe microanalysis (EPMA), which found an average Yb:Sn:Ru ratio (after Heinrich ZAF correction) of 11:39:50. Electron diffraction, using a Philips CM200 with a field-emission gun operated at 200 kV, was recorded on image plates with spot sizes less than 10 nm. Exposure times ranged from 1-3 s (condenser aperture 10 μm) for the diffraction patterns used for structure refinement and 15~30 s (condenser aperture 70 μm) for the Convergent Beam Electron Diffraction (CBED) pattern. The least squares refinement package MSLS[14] was employed to analyze the single crystal nano-electron diffraction data. This software, explicitly taking dynamical electron diffraction into account, has proved able to determine crystal structures with the same accuracy as single-crystal x-ray diffraction. These measurements revealed a cubic cell, consistent with space group Pm-3m, and a lattice constant $a \cong 3.2$ Å.

The [110] zone axis electron diffraction pattern for $Yb_{0.24}Sn_{0.76}Ru$ is shown in Fig. 1. Diffraction patterns were taken from many fragments of the specimen and showed no evidence for the presence of the $a \cong 6.4$ Å cell expected for a face-centered cubic Heusler compound with the formula $YbRu_2Sn$. If this phase were present, even at the nanoscale, it would be detectable due to the observation of a 311 reflection in the position indicated by the arrow in the Figure 1 (ref. 10).

About 30 grams of the crystals were crushed, ground and characterized by powder x-ray diffraction (XRD) and neutron powder diffraction (NPD) analysis. Powder x-ray patterns were obtained from a Bruker D8 diffractometer with Cu Kα radiation and a graphite-diffracted beam monochromator; whereas, NPD patterns were obtained on the High Intensity Powder Diffractometer (HIPD) at the Lujan Neutron Scattering Center at the Los Alamos Neutron Science Center (LANSCE) and on the high-resolution diffractometer (BT-1) at the NIST Center for Neutron Research (NCNR). The neutron powder diffraction pattern is shown on Fig. 2 a. Quantitative analysis of joint XRD and NPD data confirmed the CsCl structure, space group Pm-3m, with Ru on the 1a site (0, 0, 0) and a (Yb, Sn) mixture on the 1b site (½, ½, ½). The refined composition, $Yb_{0.24}Sn_{0.76}Ru$, is in agreement with electron microanalysis. Figure 2a shows the excellent agreement between the structural model and the diffraction data, and Table 1 gives the cell parameters.

A Scienta SES4000 spectrometer operating on the PGM-A beamline at the Synchrotron Radiation Center (SRC) in Stoughton, Wisconsin, was used for photoemission experiments. Single crystals of $Yb_{0.24}Sn_{0.76}Ru$ and $YbCu_2Si_2$ were cooled to 20 K and cleaved in vacuum of $5\times10^{-11}$ Torr. No oxide peaks were found in the spectra, which indicated clean surfaces at measurement time. In order to maximize the 4$f$ signal, valence band photoemission measurements were performed at a photon energy of 102 eV, with overall energy resolution of 20 meV.

XANES data were collected on Beamline 11-2 at the Stanford Synchrotron Radiation Laboratory (SSRL), using a Si(220) double-crystal monochomator (ϕ=90°) detuned by about 50% to reduce harmonic contamination. A sample of $Yb_{0.24}Sn_{0.76}Ru$ and of $LuRu_2Sn$ (nominal) were ground and passed through a 30-μm sieve. The resulting powder was brushed onto adhesive tape, which was then cut into strips and stacked to achieve a change in the rare-earth $L_3$ absorption across the edge corresponding to about 0.5 absorption lengths. The samples were then loaded into a LHe-flow cryostat, and absorption data were collected in transmission mode. Reported data have been pre-edge subtracted to isolate the contribution from the Yb or Lu absorption. A sample of $Yb_2O_3$ was measured simultaneously, and the reported incident energy is calibrated such that the first inflection point of that spectrum is at 8943.0 eV.

DC magnetization/susceptibility was measured with a commercial SQUID magnetometer (Quantum Design). Heat capacity and AC electrical resistivity (60 Hz) were determined in a PPMS, Quantum Design, Model 6000 system. For these heat capacity measurements, a standard relaxation calorimetry method was used. The specific heat of several single crystals (total mass 3.24 mg) also was measured to lower temperatures in a dilution refrigerator with a 9 T superconducting magnet by means of a quasi-adiabatic heat pulse method with a $RuO_2$ thermometer. The addenda contribution was subtracted. For the resistivity measurements, we used a standard four-probe technique, with two platinum wires spot-welded on one face of the crystal and two wires connected with silver paste to the opposite faces (current wires).

Inelastic neutron scattering (INS) measurements were performed on a 30-g powder sample on the High-Resolution Chopper Spectrometer (Pharos) at the Lujan center, Los Alamos Neutron Center (LANSCE), at Los Alamos National Laboratory. Assuming that the phonon scattering scales with momentum transfer as $Q^2$ and that the magnetic scattering scales with the Q-dependence of the 4*f* form factor, we subtracted the nonmagnetic component by scaling the high-Q data to the low-Q data to obtain the magnetic scattering function $S_{mag}$ [2,15,16].

**Results and discussion**

To study the Yb valence in $Yb_{0.24}Sn_{0.76}Ru$, two methods were employed, the first being valence band photoemission (PES). An $YbCu_2Si_2$ sample was measured immediately prior to measuring $Yb_{0.24}Sn_{0.76}Ru$ to establish a baseline for comparison. Significant differences are seen in the valence band electronic structure between the two samples (Figure 3). A bulk divalent Yb contribution is clearly present in the $YbCu_2Si_2$ spectrum as a pair of narrow spin-orbit split peaks. A corresponding pair of broader surface-related divalent peaks is seen in both samples at higher binding energies [17][18]. Bulk-related $Yb^{2+}$ peaks are completely absent in the spectrum for $Yb_{0.24}Sn_{0.76}Ru$. In both samples, the $Yb^{3+}$ manifold is found at higher binding energies. The lack of a divalent Yb component in the $Yb_{0.24}Sn_{0.76}Ru$ PES spectrum is in agreement with the XANES $L_3$ near-edge absorption measurement, where no sign of a divalent peak is seen (Fig. 4). Among many other examples, a similar study of $YbCu_2Si_2$, which has an Yb valence of 2.8+ reveals a clear divalent peak in the XANES [19]. An Yb valence of 3+ ($Yb^{+3}$) is therefore derived from the lack of an $Yb^{2+}$ signal in the photoemission and XANES results for $Yb_{0.24}Sn_{0.76}Ru$.

The inverse magnetic susceptibility $1/\chi$ is shown in Figure 5 a. No obvious long-range magnetic order is observed for T > 1.8 K. The high temperature (100-300 K) Curie-Weiss fit $\chi=C/(T - \theta)$ yields an effective moment of 4.4 $\mu_B$ mol-$Yb^{-1}$. This value is close to the value for the free trivalent Yb ion (4.54 $\mu_B$) whose presence was confirmed by the PES and XANES experiments. The negative value of Weiss temperature ($\theta$ = -18 K) obtained in these fits suggests the presence of antiferromagnetic (AF) correlations. In the temperature range 2-15 K, a Curie-Weiss fit gives a much smaller effective moment $\mu_{eff.}$ = 3.2 $\mu_B$ mol-$Yb^{-1}$, which, as discussed below, is due to crystal-electric-field (CEF) splitting of the $J=7/2$ manifold.

In the CsCl structure, the Yb ions are subject to a CEF of cubic symmetry [20]. In the cubic CEF, the eight-fold degenerate $4f^{13}$, $J=7/2$ state splits into two doublets ($\Gamma_6$ and $\Gamma_7$) and a $\Gamma_8$ quartet [21] with the wave functions:

$\Gamma_6 = 0.6455|\pm 7/2\rangle + 0.7638|\mp 1/2\rangle$

$\Gamma_7 = 0.866|\pm 5/2\rangle + 0.5|\mp 3/2\rangle$

$\Gamma_8 = 0.7638|\pm 7/2\rangle + 0.6455|\mp 1/2\rangle$

$0.5|\pm 5/2\rangle + 0.866|\mp 3/2\rangle$

For these wave functions, the excitation between the $\Gamma_6$ and $\Gamma_7$ states is prohibited in inelastic neutron scattering by a selection rule. Hence, if the ground state is either the $\Gamma_6$ or $\Gamma_7$ doublet, only one excitation can be observed in the INS at low temperature; whereas, if the ground state is the $\Gamma_8$ quartet, two excitations are expected.

To determine the crystal-field scheme, we measured the INS spectra using different incident energies $E_i$ to increase the dynamic range. At T = 9 K, we used $E_i$ = 35, 80, 120 and 200 meV. (The 120 and 200 meV spectra are not shown here; no magnetic scattering was observed at these higher energies.) At T = 150 K, we measured at $E_i$ = 35, 60, and 120 meV. From Fig. 6 a, it can be seen that the primary excitation occurs at 20 meV.

Over a broad energy transfer range up to 180 meV, no other peak is observed. (A possible exception is the very weak excitation seen near 35 meV; however, this does not decrease significantly with increasing Q, and may be an artifact of the phonon subtraction method.) Consistent with the selection rules, this suggests that the ground state is either the $\Gamma_6$ or $\Gamma_7$ doublet. A doublet ground state also is expected based on the magnetic entropy value of Rln2 generated up to 10 K, as discussed further below.

To better determine the ground state, we consider the low temperature magnetic susceptibility. The low temperature Curie-Weiss behavior (Fig. 5 a, inset) exhibits a Curie constant C = 1.28 emu K mol-Yb $^{-1}$ corresponding to a moment of 3.2 $\mu_B$. From the wave functions for the $\Gamma_6$ and $\Gamma_7$ states, we can calculate the matrix element of $<J_z>$. We absorb the CEF physics into the g-factor of an effective spin-1/2 doublet, where $g_{eff}$ =16$<J_z>$/7 and the low temperature Curie constant is $C_{eff}$=N$(g_{eff}\mu_B)^2$(1/2)(3/2)/3$k_B$ [22]. The calculated results are shown in Table II, where it can be seen that the $\Gamma_7$ doublet state gives a Curie constant $C_{eff}^{cal}$ value closer to the value 1.28 emu K mol-Yb $^{-1}$ measured at low temperature than the value 0.668 emu-K mol-Yb$^{-1}$ calculated for the $\Gamma_6$ doublet. This establishes that the $\Gamma_7$ doublet is the ground state and the excitation observed at 20 meV is thus believed to be the excitation from the $\Gamma_7$ doublet ground state to the $\Gamma_8$ quartet excited state. Though this assignment leaves open the question of the energy of the $\Gamma_6$ doublet, the CEF scheme is similar to that proposed [23] for the cubic YbPd$_2$Sn compound, where the $\Gamma_7 \rightarrow \Gamma_8$ transition is at 4 meV and the $\Gamma_7 \rightarrow \Gamma_6$ transition is at higher energy.

We note that at temperatures high enough to populate the $\Gamma_8$ state, an excitation between the $\Gamma_8$ and $\Gamma_6$ excited states is expected. However, as is seen from Figure 6 b, no obvious additional peak is observed. There are at least two possibilities for the lack of the $\Gamma_6 \rightarrow \Gamma_8$ excitation: 1) The $\Gamma_6$ doublet is very close in energy to the $\Gamma_8$ quartet so that the $\Gamma_8 \rightarrow \Gamma_6$ excitation is at low energy, falling in the elastic tail of the neutron measurements. 2) The $\Gamma_6$ excited level is 20 meV higher (lower) in energy than the $\Gamma_8$ excitation so that the $\Gamma_8 \rightarrow \Gamma_6$ ($\Gamma_6 \rightarrow \Gamma_8$) excitation occurs near 20 meV.

In the inset of Figure 5 b, we exhibit the inverse susceptibility 1/$\chi_{CEF}$ calculated for the $\Gamma_8$ energy $E_8$ = 20 meV and for the $\Gamma_6$ energy $E_6$ = 1, 17 meV (curves calculated for $E_6$ = 20 and 23 meV are very similar) and 40 meV. These results show that $E_6 = 1$ meV gives results closest to the measured values. However, for a $\Gamma_6$ state this close in energy to the $\Gamma_7$ state, the ground state would effectively be a quasi-quartet, which, as discussed below, is ruled out by the observation of Rln2 entropy generated up to 10 K. By adjusting the $\Gamma_6$ energy, we find that the value $E_6 = 32$ meV gives results quite close to the measured susceptibility (Fig. 5b). Hence, the assignment $E_6 = 32$ meV appears to give the best fits to the susceptibility. (If it is not an artifact of the phonon subtraction, the very small peak seen near this energy in the spectra of Fig. 6 might then reflect an admixture of the $\Gamma_8$ into the $\Gamma_7$ ground state due to a weak breaking of cubic site symmetry from alloy disorder.) However, this assignment leaves open the question as to why no $\Gamma_6 \rightarrow \Gamma_8$ excitation (for this case at 12 meV) is observed at high temperature.

Figure 7 shows the temperature dependence of the resistivity of $Yb_{0.24}Sn_{0.76}Ru$. The data exhibit a positive derivative $d\rho/dT$ and a resistance ratio (RRR) of 1.3. Though the room temperature resistivity is low, typical of a good metal, the residual resistivity is high, as might be expected from the 1b-site disorder in $Yb_{0.24}Sn_{0.76}Ru$. The red line through the data represents a fit which combines a Bloch-Grüneisen resistivity $\rho_{BG}$ together with a parallel resistor $\rho_p$[24]:

$$\rho(T)^{-1} = \rho_P^{-1} + (\rho_0 + \rho_{BG})^{-1}$$

$$\rho_{BG} = 4R\Theta \left(\frac{T}{\Theta}\right)^5 \int_0^{\Theta/T} \frac{x^5}{(\exp(x)-1)(1-\exp(-x))} dx.$$

The fit gives $\Theta = 185$ K which is very close to the Debye temperature obtained from specific heat measurement (see below). The main point here is that the resistivity is quite similar to that seen in materials where the scattering is primarily due to phonons, with no obvious contribution from magnetic scattering or the Kondo effect.

The inset of Fig. 8a presents measurements of the specific heat ($C_p$) versus temperature up to 18 K. Measurements to higher temperature show that the specific heat reaches the expected Dulong-Petit ($3nR \approx 50$ J mol$^{-1}$ K$^{-1}$) value at around 200 K. In order to calculate the magnetic contribution to the heat capacity, the electronic ($C_{el}$) and lattice ($C_{ph}$) specific heat should be subtracted from $C_p$. The usual way to determine $C_{el}$ and $C_{ph}$ is by measuring a "non-magnetic" compound which possesses the same crystal structure. However, in spite of our efforts, we were not able to grow a sample of $(Lu_{0.26}Sn_{0.74})Ru$ or LuRu of sufficient quality to obtain trustworthy specific heat data. As an alternative approach we plot specific heat coefficient C/T as a function of $T^2$ for $Yb_{0.26}Sn_{0.76}Ru$ in the inset of Fig. 8 a. The straight red line through the data is the best fit to the formula $C_p = \gamma T + \beta T^3$ in the 8-18 K temperature range. Here $\gamma T$ is the electronic contribution to the heat capacity ($C_{el}$) and $\beta T^3$ is the acoustic phonon contribution in the low-temperature limit of the Debye model. The values of $\gamma$ and $\beta$ determined from the fit are 3 mJ mol-Yb$^{-1}$ K$^{-2}$, and 0.55 mJ mol-Yb$^{-1}$ K$^{-4}$ respectively. Using the relation $\theta_D = (12\pi^4 Nk_B/5\beta)^{1/3}$, where N is the number of atoms in the unit cell and $k_B$ is the Boltzmann constant, the Debye temperature was estimated to be 191 K. This is similar to the Debye temperature 165 K estimated for $YbPd_2Sn$ [23]. Finally the magnetic specific heat ($C_{mag}$) was calculated by subtracting these contributions, i.e. $C_{mag} = C_p - \gamma T - \beta T^3$.

The magnetic specific heat for $Yb_{0.24}Sn_{0.76}Ru$ is plotted in Figs. 8 a, b and c. In the absence of an applied magnetic field, $C_{mag}/T$ increases rapidly with decreasing temperature below 3 K, reaching 9 J mol-Yb$^{-1}$ K$^{-2}$ at the lowest measured temperature 70 mK (Figure 8 a), which suggests very heavy Fermion behavior. The calculated magnetic entropy (Fig. 8 b inset) returns a value Rln2 at temperatures above 6 K, which as mentioned above rules out a quartet or quasi-quartet ground multiplet. A broad peak in $C_{mag}(T)$ is observed at 1 K, as well as a kink in $C_{mag}/T$ at T ~ 0.5 K. Though the peak is broader than expected for magnetic ordering of localized 4f electrons, the excess width might be a consequence of inhomogeneous rounding of a magnetic transition due to the large alloy disorder. To check whether the peak might arise from magnetic ordering, we performed neutron diffraction experiments at temperatures of 70 mK and 4 K. The diffraction patterns are presented in Figure 2 b. The results show no obvious difference between the diffraction intensities at these two temperatures. Hence, it is very unlikely that the anomaly at 1 K arises from long-range antiferromagnetic order.

A second possible explanation for the peak in $C_{mag}$ is that it originates from a Kondo contribution. This would be consistent with the negative Weiss temperature that is determined from the low temperature susceptibility by fitting to the form $\chi_{Cal} = 1/(1/\chi_{CEF} + \lambda)$. Such a Weiss temperature can arise from Kondo physics, where $T_K = \lambda C_{eff}$. Given $C_{eff} = 1.28$ emu K mol-Yb$^{-1}$, the value $\lambda = 2$ mol-Yb emu$^{-1}$ for the $E_6 = 32$ meV fit suggests $T_K = 2.6$ K. Because a $\Gamma_7$ doublet ground state is implied by both the magnetic susceptibility and the inelastic neutron scattering measurements, we plot the expected [25] Kondo specific heat for a doublet (effective $J = ½$ case). We use the value of $T_K = 1$ K in order to obtain the correct temperature for the peak. It can be seen from the comparison that the Kondo behavior is too broad compared to the experimental data.

A third possibility is that antiferromagnetic short-range order contributes to the specific heat anomaly. (We note that a negative Weiss temperature also can arise from antiferromagnetic correlations, where $T_N = \lambda C_{eff}$.) In this scenario, the antiferromagnetic interaction acts as an internal magnetic field $B_{int}$ which induces a Zeeman splitting of the $\Gamma_7$ doublet ground state even in zero applied field, and gives a Schottky-like anomaly in $C_{mag}$ at 1 K [22]. Aoki et al [23] have shown that in an applied field under these circumstances, the antiferromagnetic interaction results in an effective distribution of the total local field that splits the doublet. A consequence is that the Schottky anomaly observed in the specific heat in an applied field is shifted to lower temperature, is broader, and has a smaller peak magnitude (for YbPd$_2$Sn, about 20% smaller[23]) than would be the case in the absence of the AF interaction.

In Fig. 8 c, the solid lines are Schottky curves calculated for Zeeman splitting of a doublet due to a single (undistributed) local field, using the effective g-factor of the $\Gamma_7$ doublet. At each applied field, the local field $B_{loc}$ is adjusted to give the best fit to the data. At $B_{app} = 0$, this results in an internal field of 1 T. We note that the Schottky curves have been multiplied by a factor of 0.75 which, as mentioned, is a factor similar to that needed to account for specific heat measurements on YbPd$_2$Sn. In addition, the measured specific heat curves are broader than the Schottky curves, especially at low temperature. The reduction in amplitude and the broadening can arise from both the distribution of local fields that arise from the AF short range order, but also from the alloy disorder. It can be seen from Fig. 8 d that the observed peak temperatures are consistently smaller than the values expected when the antiferromagnetic interaction vanishes. Indeed, in finite applied field, the peak temperatures come very close to the value expected for a total local field of $B_{loc} = B_{app} - 1$ T. This analysis supports our contention that the specific heat anomaly comes from a Zeeman splitting of the $\Gamma_7$ doublet ground state due to a combination of antiferromagnetic fluctuations and applied external field.

The existence of antiferromagnetic short-range order is a common occurrence in paramagnetic heavy-Fermion compounds residing close to a quantum-critical point where there is a zero-temperature transition between paramagnetic and antiferromagnetically ordered ground states [12]. Under these circumstances, the AF fluctuations would coexist with Kondo fluctuations[26] and the peak in the specific heat observed for $Yb_{0.26}Sn_{0.76}Ru$ at low temperature would reflect a combination of Kondo physics and AF fluctuations. The Kondo temperature would have to be quite small – in the range 1-2 K given the small Weiss temperature and the large $C_{mag}/T$ at 70 mK.

However, the short ranged order and the large value of the specific heat coefficient observed in this compound do *not* require proximity to a quantum-critical point. The absence of long-range order is mandated by the fact that the site occupancy of Yb is below the percolation threshold (x = 0.311 for a cubic lattice [27]). For such a dilute system, the size of clusters of directly interacting Yb atoms is small, which limits the coherence length to a finite value. Isolated Yb impurities can also exist, and can hybridize with the conduction electrons, so Kondo physics is not ruled out, but the primary contribution to the specific heat peak comes from the onset of the AF correlations.

**Conclusions**

Single crystals of $Yb_{0.24}Sn_{0.76}Ru$ were synthesized and shown to form in the CsCl structure. The magnetic susceptibility, specific heat, valence band photoemission, $L_3$ x-ray absorption, and inelastic neutron scattering spectra indicated that the Yb is trivalent in this compound and is subject to crystalline electric fields in the cubic site symmetry. If, as is common, the minimum temperature of our measurements had been only 1-2 K, we might have concluded from the very large C/T and the non-quadratic temperature dependence of the resistivity that $Yb_{0.24}Sn_{0.76}Ru$ is at or near a quantum-critical point. By extending specific heat measurements to temperatures well below 1 K and determining the field response, however, we showed that this does not have to be the case. Instead, the broad suite of experiments employed in this study collectively lead to the conclusion that all properties of this material can be understood primarily to arise from conventional local moment physics that is strongly influenced at low temperatures by the dilution and disorder of Yb atoms. The resulting antiferromagnetic short-ranged order (SRO) occurs because the Yb concentration is close to, though smaller than, the percolation limit. It would be interesting to study the evolution of the behavior as a function of Yb concentration from the very dilute limit, through the SRO regime, into the regime of AF long-range order. In addition, since the possibility of new states or new examples of unconventional superconductivity in strongly correlated Yb-based compounds merits further exploration, it would be interesting to synthesize a chemically ordered variant of $Yb_{0.24}Sn_{0.76}Ru$.

**Acknowledgements**

Work by the U C Irvine group was supported by the US Department of Energy (DOE) under grant DE-FG02-03ER46036. Work at Los Alamos was performed under the auspices of the US Department of Energy, Office of Basic Energy Sciences, Division of Materials Science and Engineering. Work at ORNL was sponsored by the Laboratory Directed Research and Development Program of Oak Ridge National Laboratory, managed by UT-Battelle, LLC, for the US DOE. Work at LBNL was supported by the Director, Office of Science, OBES, of the US DOE under Contract No. DE-AC02-05CH11231. This work is based in part upon research conducted at the Synchrotron Radiation Center, University of Wisconsin-Madison, which is supported by the National Science Foundation under Award No. DMR-0537588. This work has benefited from the use of HIPD at the Lujan Center at the Los Alamos Neutron Science Center, funded by DOE Office of Basic Energy Sciences. Los Alamos National Laboratory is operated by Los Alamos National Security LLC under DOE Contract DE-AC52-06NA25396. We acknowledge the support of the National Institute of Standards and Technology, U.S. Department of Commerce, in providing the neutron research facilities used in this work. The authors acknowledge financial support from the European Union under the Framework 6 program under a contract for an Integrated Infrastructure Initiative; reference026019ESTEEM.

Table I. Structural parameters for $Yb_{0.24}Sn_{0.76}Ru$ determined from simultaneous refinement of x-ray and neutron diffraction data at 298 K. Space group Pm-3m (s.g. #221), a = 3.21729(8) Å. Calculated density 11.6 g cm$^{-3}$. The $U_{iso}$ are the thermal vibration parameters in $10^{-2}$ Å$^2$. Figures of merit: goodness of fit $\chi^2$ = 12.56 for 96 variables, residual on structure factors $R(F^2)$ = 8.41%.

**$Yb_{0.24}Sn_{0.76}Ru$ at 298 K**

| Atom | $U_{iso}$ | Position | Occupancy |
|---|---|---|---|
| Ru | 0.576(10) | 1a (0,0,0) | 1.000 |
| Yb | 0.336(8) | 1b (0.5, 0.5, 0.5) | 0.236(6) |
| Sn | 0.336(8) | 1b (0.5, 0.5, 0.5) | 0.764(6) |

Table II. The calculated effective low temperature Curie constant for the two different pseudo spin 1/2 doublet ground states possible in the CsCl structure.

| | $<J_z>$ | $g_{eff}$ | $C_{eff}^{cal}$ (emu K mol-Yb$^{-1}$) |
|---|---|---|---|
| $\Gamma_6$ | 1.167 | 2.667 | 0.668 |
| $\Gamma_7$ | 1.5 | 3.428 | 1.104 |

**Captions**

Fig.1

Nano-electron diffraction pattern of $Yb_{0.24}Sn_{0.76}Ru$ in the [110] zone, indexed on the primitive cubic $a = 3.21$ Å cell. For a Heusler phase crystal structure, which is a supercell of this structure, $a$ would be doubled, the reciprocal lattice would be scaled by 1/2 (i.e. the reflections marked 001 and 110 would be the 002H and 220H reflections respectively), and FCC systematic absences would be observed. In FCC, the expected presence of reflections for h,k,l all odd would yield spots at the positions 111H, 113H (marked with arrow), 331H etc., which are not seen. The absence of these reflections, even at the nanometer length scale, indicates that the crystal structure of $Yb_{0.24}Sn_{0.76}Ru$ is primitive cubic.

Fig. 2 (color online)

a) Rietveld refinement of room temperature powder neutron diffraction data for $Yb_{0.24}Sn_{0.76}Ru$ obtained on the High Intensity Powder Diffractometer (HIPD) at the Lujan center, LANSCE. Upper part: open blue circles - observed data, red solid line - calculated intensities. The blue tick marks correspond to $Yb_{0.24}Sn_{0.76}Ru$, green, red and black sets referring to the impurities Ru (4.0%), $Yb_2O_3$ (1.1%), and Pb (0.7%) respectively. The lower part shows on the same scale the differences between the observed and calculated pattern.

b) Low temperature (open circles – 4 K, solid line – 70 mK) diffraction patterns obtained on High Resolution Powder Diffractometer (BT1) at NIST. Error bars represent ± 1σ. The lower part shows on the same scale the difference between diffraction patterns obtained at two different temperatures.

Fig. 3 (color online)

Valence band photoemission data for $Yb_{0.24}Sn_{0.76}Ru$, showing bulk and surface $Yb^{2+}$ and bulk $Yb^{3+}$ contributions. The bulk $Yb^{2+}$ features seen in mixed-valent $YbCu_2Si_2$ are absent from the spectrum for $Yb_{0.24}Sn_{0.76}Ru$, indicating trivalent Yb in that compound.

Fig. 4 (color online)

a) The $L_3$ x-ray absorption spectrum for $Yb_{0.24}Sn_{0.76}Ru$ at 50 and 300 K and

b) a comparison of the derivative of the $L_3$ spectrum for $Yb_{0.24}Sn_{0.76}Ru$ (50 K) and $LuRu_2Sn$ (nominal composition, 240 K). Also shown for comparison in each panel is a Yb $L_3$-edge spectrum collected at 20 K for the intermediate valence compound $Yb_{0.1}Lu_{0.9}Al_3$ (68% trivalent Yb) from [28], shifted in energy so that the main peak position agrees with $Yb_{0.24}Sn_{0.76}Ru$. A divalent component would manifest as a peak in the white line near 8935 eV, or in the derivative near 8933 eV, as shown in the $Yb_{0.1}Lu_{0.9}Al_3$ spectrum. No divalent component is seen in these spectra, implying that the Yb is trivalent in $Yb_{0.24}Sn_{0.76}Ru$.

Fig. 5 (color online)

a) Inverse magnetic susceptibility $1/\chi$ of $Yb_{0.24}Sn_{0.76}Ru$. The solid line is the high temperature Curie-Weiss fit. The inset compares the measured $1/\chi$ at low temperature to a Curie-Weiss fit.

b) The susceptibility and inverse susceptibility (inset) compared to calculations based on the crystal field model given in the text, with various values for the $\Gamma_6$ energy $E_6$, and for a Weiss field $\lambda$. The best fit is for $E_6 = 32$ meV and $\lambda = 2$ mol-Yb emu$^{-1}$.

Fig. 6 (color online)

Magnetic contribution $S_{mag}$ to the inelastic neutron scattering (INS) spectra of $(Yb_{0.24}Sn_{0.76})Ru$. The data were taken on Pharos at T = 9 K (a) and 150 K (b) with different incident energies to increase the dynamic range of the INS spectrum. One obvious excitation is seen at an energy transfer $\Delta E = 20$ meV, corresponding to the $\Gamma_7/\Gamma_8$ crystal field excitation.

Fig. 7 (color online)

Temperature dependence of the resistivity of $Yb_{0.24}Sn_{0.76}Ru$. The resistivity below 20K was measured under an applied magnetic field of 1T to remove the superconducting signal from the remaining traces of Pb flux. The red line is a fit which combines a Bloch-Grüneisen resistivity $\rho_{BG}$ together with parallel resistor $\rho_p$, as described in the text.

Fig. 8 (color online)

(a) Sommerfeld coefficient of the magnetic specific heat $C_{mag}/T$. The inset shows $C_p/T$ versus $T^2$ for the measured total specific heat $C_p$; the red line is the sum of the phonon and electronic contributions.

(b) Magnetic contribution $C_{mag}$ to the specific heat compared to a curve calculated for a Kondo doublet. The inset is the magnetic entropy.

(c) $C_{mag}$ in different applied magnetic fields. The solid lines are Schottky specific heat curves (multiplied by a factor 0.75) where the local field is adjusted to give the best fit to the data.

(d) The observed temperature of the specific heat Schottky anomaly compared to the value expected when there is no antiferromagnetic interaction, i.e. when $B_{loc}=B_{app}$ (blue line), and also to the value expected when the total local field is $B_{app} - 1$ T (red diamonds).

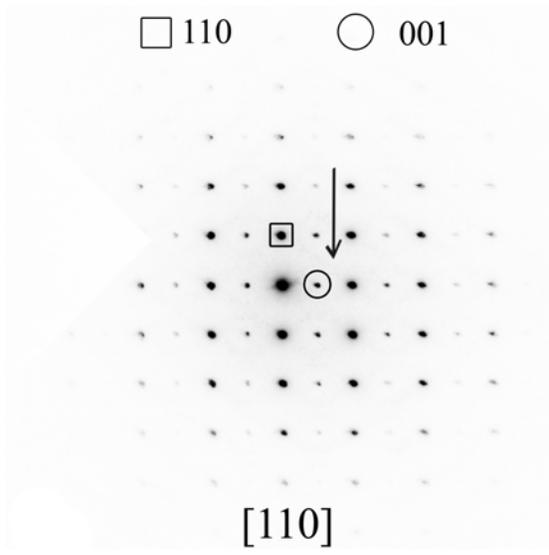

Figure 1.

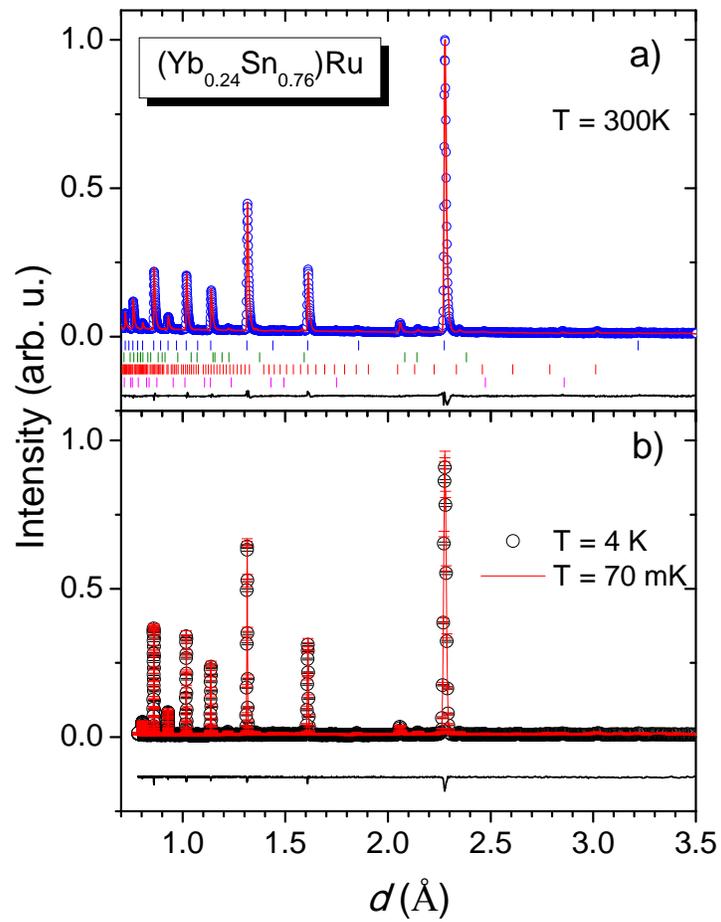

Figure 2.

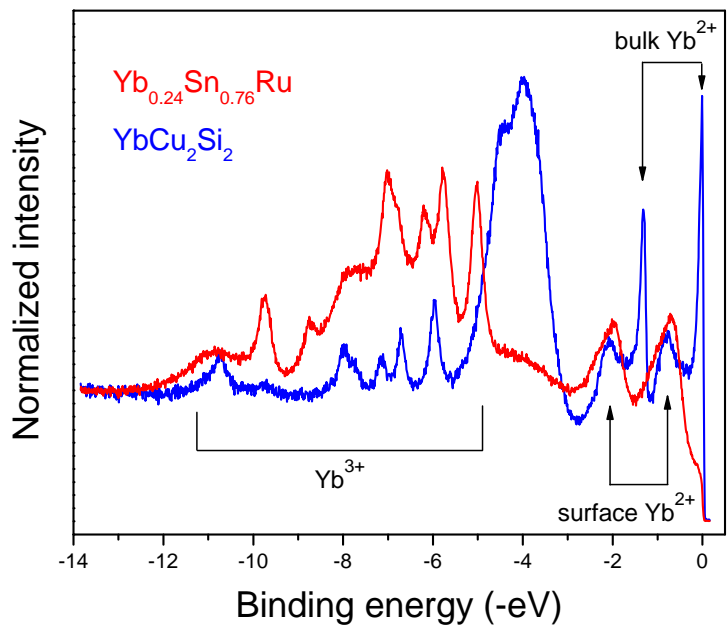

Figure 3.

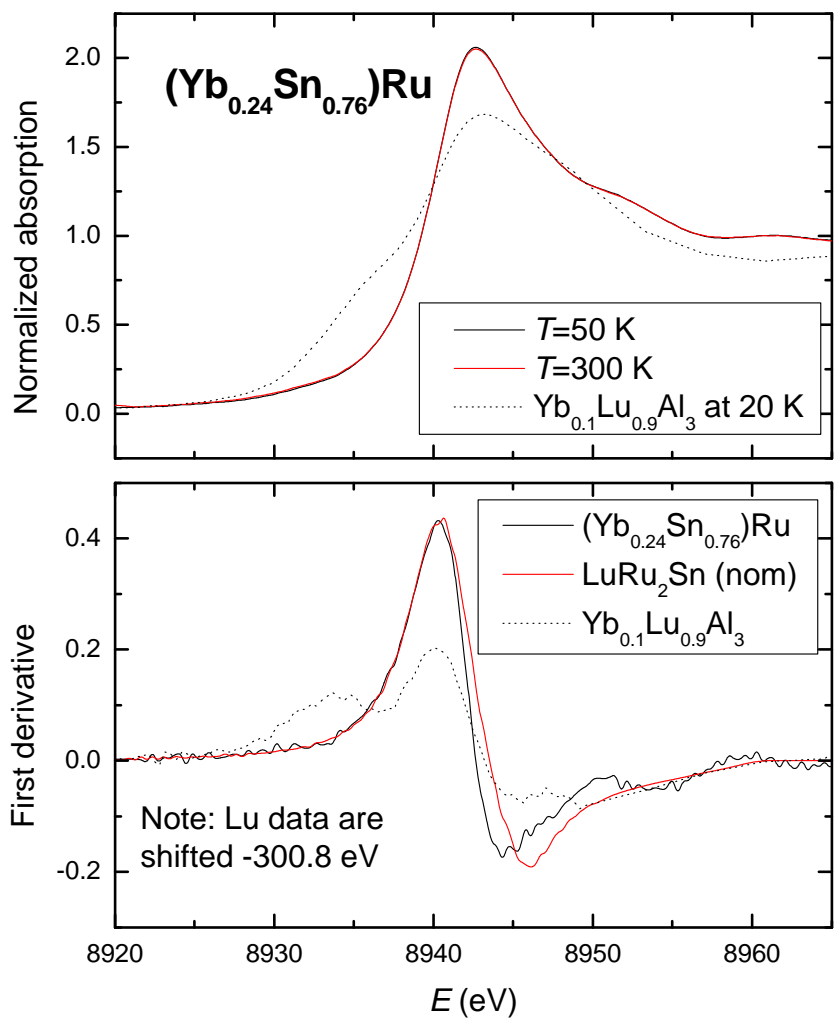

Figure 4.

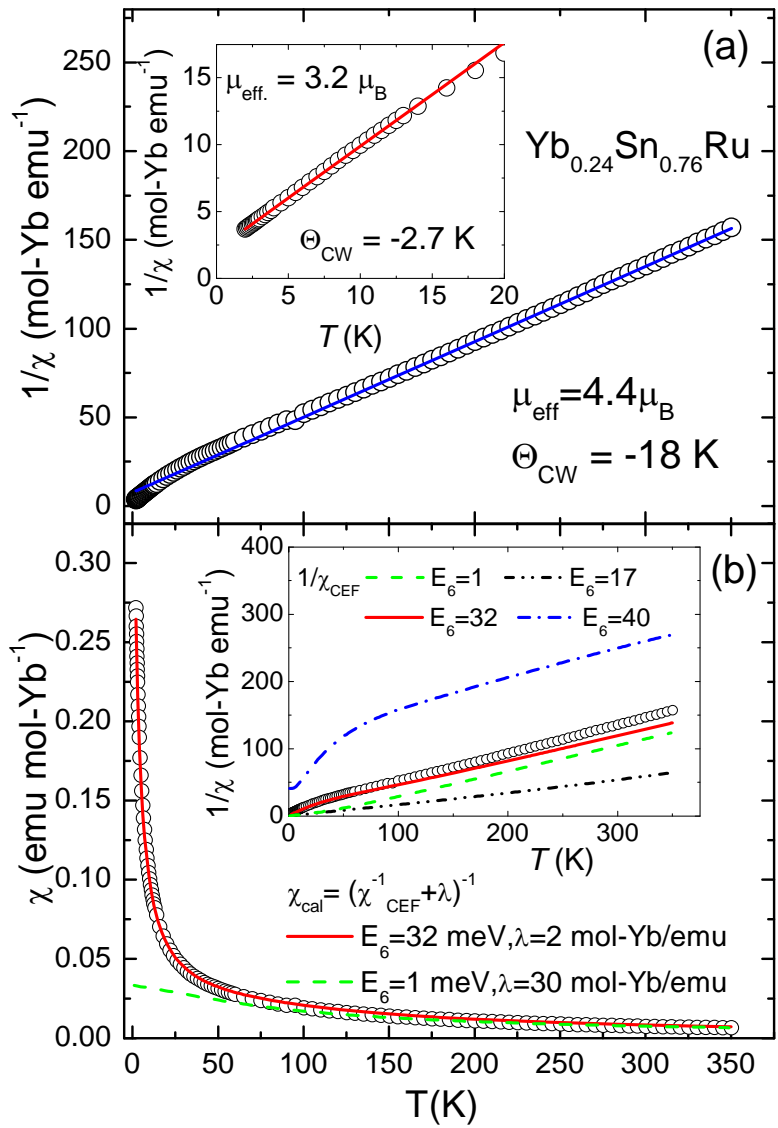

Figure 5

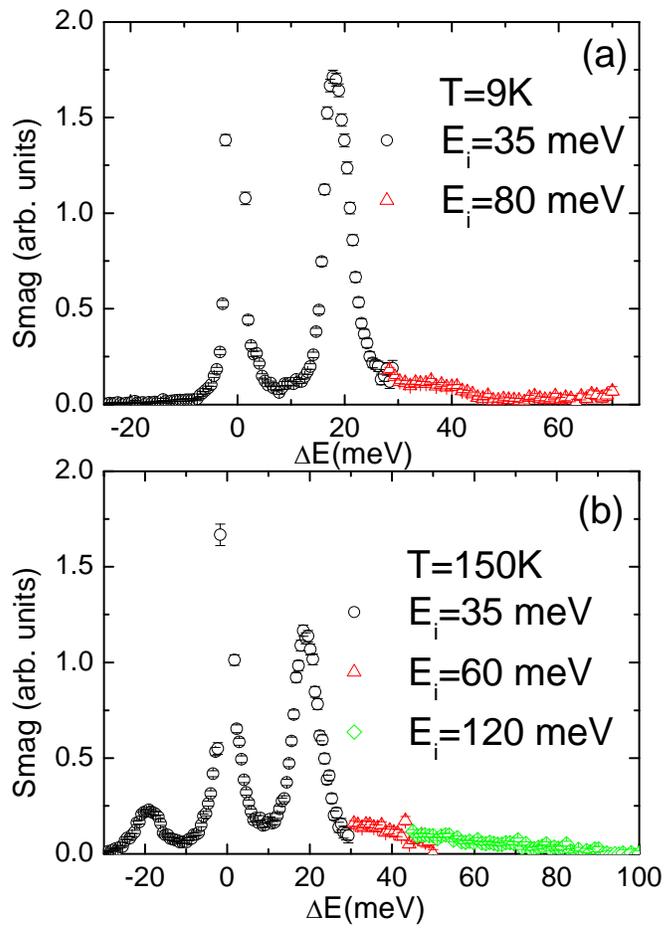

Figure 6.

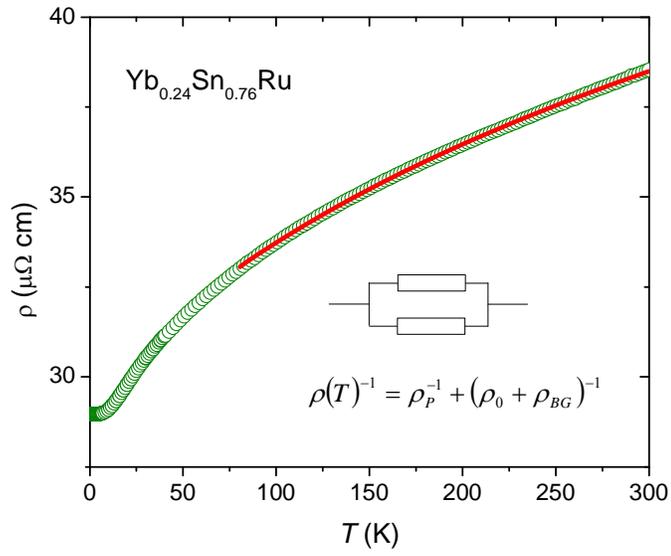

Figure 7.

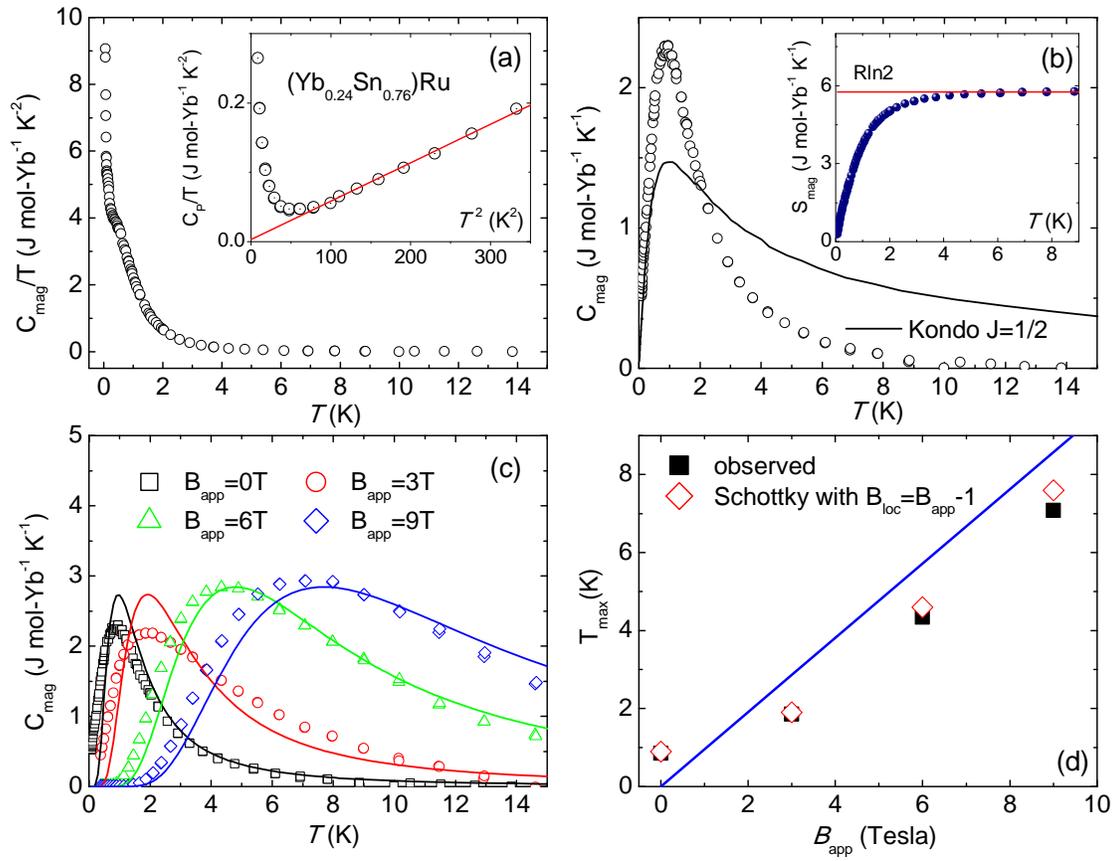

Figure 8.